\documentclass[reprint,prl,twocolumn,superscriptaddress,nofootinbib,floatfix]{revtex4-1}
\usepackage{graphicx,bm,color,amsmath,amssymb}
\newcommand{\non}{\nonumber}
\newcommand{\ba}{\begin{eqnarray}}
\newcommand{\ea}{\end{eqnarray}}
\begin{document}

\title{Conditions for defocusing around more general metrics in Infinite Derivative Gravity}
\author{James Edholm}
\affiliation{Lancaster University}

\begin{abstract}Infinite Derivative Gravity is able to resolve the Big Bang
curvature singularity present in general relativity by using a simplifying
ansatz. We show that it 
can also avoid the Hawking-Penrose singularity, by allowing defocusing of
null rays through the Raychaudhuri equation. This 
occurs not only in the minimal case where we ignore the matter contribution,
but also in the case where matter plays a key role.

We investigate the conditions for defocusing for the general case where this
ansatz applies and also for more specific metrics, including a general Friedmann-Robertson-Walker
(FRW)
metric and three specific choices of the scale factor which produce a bouncing
FRW universe.
\end{abstract}
\maketitle
\section{Introduction}
The theory of General Relativity (GR) has been shown to describe gravity very accurately through a huge range of
experimental tests over the past century~\cite{Will:2014kxa}. However, it suffers from problems, generating both black hole and cosmological singularities~\cite{Hawking:1973uf}.

Previous attempts to solve this problem include $f(R)$ gravity and higher derivative gravity. Higher derivative gravity suffers from the Ostrogradsky instability
 which produces ghosts~\cite{Barnaby:2007ve}, which are physical excitations with negative kinetic energy~\cite{VanNieuwenhuizen:1973fi}.
Infinite Derivative Gravity (IDG) solves this problem by adding an infinite sum of the d'Alembertian operator $\Box=g^{\mu\nu} \nabla_\mu \nabla_\nu$, 
acting on the curvature. 
There is no highest derivative operator, so the Ostrogradsky instability does not apply. We must show that there are no ghosts in other ways. 
It has been shown that if the modification to the propagator contains at most a single pole then the problem of ghosts is avoided~\cite{Biswas:2005qr,Biswas:2010zk}.

Infinite derivative actions, which are used in string theory~\cite{Tseytlin:1995uq}, 
were first applied to gravity by Biswas, Gerwick, Kovisto and Mazumdar~\cite{Biswas:2011ar}. 
IDG  has been investigated around flat Minkowksi backgrounds~\cite{Koshelev:2017bxd}, (Anti) de Sitter 
backgrounds~\cite{Biswas:2016etb,Biswas:2016egy,Conroy:2015nva,Edholm:2018wjh}, a rotating metric~\cite{Cornell:2017irh} and the Schwarzschild black hole solution~\cite{Calcagni:2017sov}.

It is possible to find the gravitational entropy of a black hole within IDG~\cite{Conroy:2015wfa,Conroy:2015nva}. 
The propagator can also be found for this theory~\cite{Tomboulis,Siegel:2003vt,Biswas:2005qr,Biswas:2011ar,Biswas:2016etb,Biswas:2013kla,Buoninfante:2016iuf}, 
and it has been shown 
that one can curtail the divergences of 1 and 2 loop diagrams~\cite{Talaganis:2014ida,Talaganis:2015wva}, while other work has investigated the
UV finiteness~\cite{Talaganis:2017tnr}. 

Within the Arnowitt-Deser Misner (ADM) decomposition, one can find the boundary
terms of the theory~\cite{Teimouri:2016ulk}.
IDG can be thought of as an extension to Starobinsky inflation~\cite{Briscese:2012ys,Biswas:2012bp,
Koshelev:2016xqb,Craps:2014wga,Koshelev:2017tvv},
allowing us to put a constraint on the mass scale $M$~\cite{Edholm:2016seu}. 
It is also possible to put constraints on $M$ using
the deflection of light by the Sun~\cite{Feng:2017vqd}, or by comparing the Newtonian potential 
to experimental evidence on
the strength of gravity at small distances~\cite{Edholm:2016hbt,Conroy:2017nkc}.

GR suffers from the singularity problem. Due to the Einstein equations, the Raychaudhuri equations~\cite{waldbook} 
make it impossible for null rays to defocus as long as the Null Energy Condition holds,
which by the Hawking-Penrose singularity theorem means there must be a singularity.
IDG has different equations of motion to GR which generate the possibility of avoiding this singularity,
through adding extra terms to the Einstein-Hilbert action.
The conditions necessary to allow defocusing were investigated around a Minkowski background~\cite{Conroy:2016sac,Conroy:2017nkc,Conroy:2017uds}, 
around an (Anti) de Sitter background~\cite{Edholm:2017fmw} and for an FRW metric near the bounce~\cite{Conroy:2014dja}.

\section{The Raychaudhuri condition}
The Raychaudhuri equation is a model-independent geometrical equation that tells us the expansion 
of a congruence of null rays emerging from the centre of our coordinate system,  
which have the tangent vectors $k_\mu$ where $k^\mu k_\mu=0$~\cite{waldbook,Relativiststoolkit}. 
If we imagine that the outgoing null rays form the surface of a sphere, then the expansion parameter 
$\theta=\nabla_\mu k^\mu$ describes the change in volume of that sphere. 
We would expect that for these outgoing null rays, the sphere would be expanding, 
but for a \textit{trapped surface}, the volume actually decreases and $\theta<0$. 
By the Raychaudhuri equation, $\theta$ fulfils the condition 
\ba
        \frac{d\theta}{d\tau} + \frac{1}{2} \theta^2 \leq - R_{\mu\nu} k^\mu k^\nu, 
\ea        
where $R_{\mu\nu}$ is the Ricci curvature tensor and $\tau$ is the affine parameter. 
Here we have taken the twist to be zero and we have used that the shear term is strictly 
positive to turn the Raychaudhuri equation into an inequality\footnote{The rotation term 
vanishes if we take the congruence of null
 rays to be orthogonal to a hypersurface and the shear term is strictly positive because the shear tensor is purely transverse~\cite{waldbook,Relativiststoolkit}.}. If the null rays are to defocus, then the expansion parameter must be both positive and expanding,
which implies that 
\ba \label{eq:riccidefocusingcondition}
        R_{\mu\nu}k^\mu k^\nu<0,
\ea
which we call the defocusing condition. Unless this condition is satisfied, the 
Hawking-Penrose singularity theorem says that  a singularity will be generated~\cite{Hawking:1973}.

The Null Energy Condition (NEC) says that for non-exotic matter, 
the stress-energy tensor $T_{\mu\nu}$ contracted with the tangent vectors is non-negative, i.e. 
$T_{\mu\nu} k^\mu k^\nu \geq 0$. 
Iif one inserts the Einstein equation into \eqref{eq:riccidefocusingcondition}
it can be seen that the defocusing condition is not fulfilled and therefore by the 
Hawking-Penrose singularity theorem, GR generates singularities. 
The defocusing condition has also been investigated for f(R) gravity~\cite{Carloni:2005ii}.
It was shown for perturbations around Minkowski and for a bouncing FRW solution near the bounce
that IDG could allow defocusing of null rays~\cite{Conroy:2016sac,Conroy:2017nkc,Conroy:2017uds,Conroy:2014dja}.

\section{Infinite Derivative Gravity}
We will look at the IDG action
\ba
        S = \frac{1}{2}\int d^4 x \left[M^2_P R +  R F(\Box)R-2\Lambda\right],
\ea        
where $R$ is the Ricci curvature scalar, $\Lambda$ is the cosmological constant with mass dimension 4 and $M_P$ is the Planck mass. $F(\Box)\equiv\sum_{n=0}^\infty f_n \Box^n/M^{2n}$
is the infinite sum of the d'Alembertian operator $\Box\equiv g^{\mu\nu} \nabla_\mu \nabla_\nu$, 
regulated by the mass scale $M$ and with the dimensionless coefficients $f_n$. This action produces the equations of motion~\cite{Biswas:2013cha}  
\begin{widetext}
\ba \label{eq:fulleoms}
                T_{\alpha\beta} &=& M^2_p \left(R_{\alpha\beta} -\frac{1}{2}g_{\alpha\beta} R\right) + g_{\alpha\beta} \Lambda+ 4  \left(R_{\alpha\beta} -\frac{1}{2} g_{\alpha\beta} R\right) F(\Box)R
                + g_{\alpha\beta} R F(\Box)R - 4 \left(\nabla_\alpha \nabla_\beta - g_{\alpha\beta} \Box\right) F(\Box)R \nonumber\\
                &&+\sum_{n=1}^\infty f_n \sum_{l=0}^\infty\left[ g_{\alpha\beta} \left( \left(\partial^\sigma \Box^lR \right)\partial_\sigma \Box^{n-l-1}R +\Box^l R \Box^{n-l} R\right) 
                -2  \left(\partial_\alpha \Box^lR \right)\partial_\beta \Box^{n-l-1}R\right].
\ea
\end{widetext}
where $T_{\alpha\beta}$ is the stress-energy tensor and $R_{\alpha\beta}$ is the Ricci curvature tensor. If we contract \eqref{eq:fulleoms} with tangent vectors $k^\alpha k^\beta$, where $k^\alpha k_\alpha=0$, we can find the full condition for null rays to defocus \eqref{eq:riccidefocusingcondition}, 
which allows us to avoid the Hawking-Penrose singularity\footnote{(\ref{eq:defocusingcondition}) reduces to 
the defocusing condition found for a bouncing FRW cosmology~\cite{Conroy:2014dja} 
if we take the Ricci scalar $R$ to be a function of $t$ only and consider only even powers
of $t$ (as happens at the bounce), so that $H\to 0$ 
 and take $\alpha=1/2$.}\footnote{The cosmological constant does not feature in \eqref{eq:defocusingcondition} due to the condition $k^\alpha k^\beta g_{\alpha\beta}=0$. }
 \ba \label{eq:defocusingcondition}
         \hspace{-4mm}k^\alpha k^\beta R_{\alpha\beta}
         &=&\frac{1}{M^2_p 
+ 4 F(\Box)R}\bigg[k^\beta k^\alpha T_{\alpha\beta} 
                 + 4 k^\beta k^\alpha\nabla_\alpha
\nabla_\beta  F (\Box)R \nonumber\\
&&\hspace{-4mm}+ 2 k^\beta k^\alpha \sum^\infty_{n=1} \frac{f_{n}}{M^{2n}} \sum^{n-1}_{l=0}\left(\partial_\alpha
\Box^l R\right) \partial_\beta \Box^{n-l-1} R\bigg]<0.~~~~~
\ea
\subsection{The simplifying ansatz}
In the next section we will show IDG also allows null rays to defocus for
solutions 
where the d'Alembertian operator $\Box$ acting on the Ricci scalar $R$ fulfills
the ansatz $\Box R = r_1 R + r_2$, where $r_1$ and $r_2$ are constants, as
studied in previous work~\cite{Biswas:2005qr,Biswas:2010zk,Koshelev:2012qn,Koshelev:2013lfm,Koshelev:2017tvv}.
In fact, \cite{Koshelev:2017tvv} showed that a metric of this form was the
most
general spatially flat FRW solution within IDG.  

Using this ansatz,~\cite{Biswas:2005qr,Biswas:2010zk,Koshelev:2012qn,Koshelev:2013lfm} were able to find vacuum solutions to the equations of motion (\ref{eq:fulleoms}).
These focused on generating bouncing FRW cosmologies, in particular solutions where $a(t)=\cosh(\sigma t)$ and $a(t)=e^{\frac{\lambda}{2} t^2}$. 
A simple toy model $a(t)=1+a_2t^2$ was studied as a perturbation to flat space in~\cite{Conroy:2014dja}. 
These bouncing solutions exhibit time symmetry about $t=0$ and avoid the curvature singularity at $t=0$ present in GR, known as the Big Bang singularity problem.  
We will examine the most general metric
where this ansatz is satisfied, and then look at these more specific metrics, showing that we can avoid the singularity generated by General Relativity\ through the Hawking-Penrose singularity theorem~\cite{Hawking:1973}.
 
\subsubsection{Minimum defocusing condition using the ansatz}
If we take the ansatz $\Box R=r_1 R + r_2$, then for $n>0$, $\Box^n R = r_1^n\left(R +\frac{r_2}{r_1}\right)$ and 
\ba \label{eq:defnofboxwithansatz}
        F(\Box) R=\sum_{n=0}^\infty f_n\frac{\Box^n}{M^{2n}}R= F(r_1)\left(R+\frac{r_2}{r_1}\right)-f_{0}r_2/r_1,~~~~~
\ea        
where $f_0$ is the zero-order coefficient of $F(\Box)$, i.e. the coefficient of $R^2$ in the action. 
The minimum defocusing condition \eqref{eq:defocusingcondition} 
becomes\footnote{We can return to a local theory by taking $r_1R=-r_2$, which removes the dependence on higher orders of $\Box$ in 
\eqref{eq:defnofboxwithansatz}. This can be clearly seen in the denominator of \eqref{eq:mindefocusingconditionwithansatz}. 
It is not explicitly seen in the numerator of \eqref{eq:mindefocusingconditionwithansatz} because the derivatives of $r_2$ vanish, so $r_1R=-r_2$ implies that the derivatives of $r_1 R$ also
vanish, so the higher order terms do not contribute. 
} 
\begin{widetext}
\ba \label{eq:mindefocusingconditionwithansatz}
         \frac{1}{M^2_p 
+4 F(r_1)\left(R+\frac{r_2}{r_1}\right)-4 f_{0}r_2/r_1 
}&&\bigg[ k^\beta k^\alpha T_{\alpha\beta} 
                 +4 k^\beta k^\alpha \sum_{n=0}^\infty \frac{f_{1_n}}{M^{2n}}\nabla_\alpha
\nabla_\beta \left(r^n_1 R\right) \nonumber\\
&&+2k^\beta k^\alpha \sum^\infty_{n=1}\frac{f_{n}}{M^{2n}}\sum^{n-1}_{l=0}
\partial_\alpha  \left(r^l_1 R\right)\partial_\beta \left(r_1^{n-l-1} R\right)\bigg]<0,
\ea
\end{widetext}
where we are looking at vacuum solutions, so neglect the stress-energy tensor term and have also removed the derivatives of the constants $r_1$ and $r_2$. 
We can simplify \eqref{eq:mindefocusingconditionwithansatz}. 
$r_1^{n-2}$ and the partial derivatives of $R$ have no $l$
dependence and can therefore be pulled out of the sum. 
By observing that $\sum^{n-1}_{l=0}1=n$ and  $k^\beta k^\alpha\left(\partial_\alpha R\right)\partial_\beta  R=\left(k^\alpha \partial_\alpha R\right)^2$, 
our final condition for the conditions for defocusing for any metric which fulfils $\Box R = r_1 R + r_2$ can be written as\footnote{This matches what we find if we apply the ansatz to perturbations around a Minkowski or de Sitter background when we discard the appropriate terms~\cite{Conroy:2016sac,Conroy:2017uds}.} \begin{widetext}
\ba 
         &&\frac{1}{M^2_p 
+4 F(r_1)\left(R+\frac{r_2}{r_1}\right)-4 f_{0}r_2/r_1 
}\bigg[ k^\beta k^\alpha T_{\alpha\beta} 
                 +4 \sum_{n=0}^\infty \frac{f_{1_n}}{M^{2n}}r^n_1 k^\beta k^\alpha\nabla_\alpha
\nabla_\beta  R +2\sum^\infty_{n=1}\frac{nf_{n}}{M^{2n}}r^{n-1}_1 \left(k^\alpha \partial_\alpha R\right)^2 \bigg]<0,~~~~~~~
\ea
or more succinctly as\footnote{This correctly reduces to $R+R^2$ gravity~\cite{Albareti:2012va} in the limit $r_1=r_2=0$.}
\ba \label{eq:defocusingconditionforgenericf}
        &&\frac{1}{M^2_P +4 F( r_1) (R+\frac{r_2}{r_1}) -4f_0\frac{r_2}{r_1}} \bigg[ k^\beta k^\alpha T_{\alpha\beta} 
                 +4 F\left(r_1\right) k^\beta k^\alpha \nabla_\alpha
\nabla_\beta R +2 F'(r_1)\left(k^\alpha \partial_\alpha R\right)^2 
        \bigg]<0,~~~~~~~
\ea
\end{widetext}
where $F'(\Box)$ is defined as
\ba
        F'(\Box)\equiv  \sum^\infty_{n=1}\frac{nf_n}{M^{2n}}\Box^{n-1}.        
\ea

\section{The FRW metric}
FRW metrics describe universes that are time-dependent, homogenous and isotropic. These metrics can suffer from curvature singularities at $t=0$.
We will investigate some bouncing universes that can avoid these curvature singularities and see whether they can avoid the Hawking-Penrose singularity.

 An FRW metric in spherical coordinates takes the form
\ba
        ds^2=-dt^2 +a^2(t)\left(\frac{dr^2}{1-\kappa r^2}+r^2d\Omega^2\right),~~~~
\ea        
where $a(t)$ is the scale factor of the universe and $\kappa$ is the spatial curvature. 

Using the condition on the tangent vectors $k^\mu k_\mu=0$ gives (noting that $R=R(t)$ even if the spatial curvature $\kappa$ is non-zero) 
\ba
        k^\beta k^\alpha \nabla_\alpha
\nabla_\beta R(t) 
= (k^0)^2\left( \partial_0^2 - H \partial_0
\right)R(t).
\ea        
We can also simplify the defocusing condition \eqref{eq:defocusingconditionforgenericf} by noting that for an FRW metric, the background d'Alembertian acting on a time-dependent 
scalar quantity $S(t)$ is given by $\Box S(t)= -\partial_0^2 S(t) -3H\partial_0 S(t)$, so $\partial_0^2 S(t)
-H\partial_0 S(t)=-\Box S(t)-4H\partial_0 S(t)$, where $H$ is the Hubble parameter. Therefore by comparing this to the ansatz, we find $\partial_0^2 R(t)
-H\partial_0 R(t)=-r_1 R(t)-4H\partial_0 R(t)$. We can divide by $(k^0)^2$ as it is strictly positive and use that $H\approx 0$ near the bounce point. 
Finally, note that for FRW, the null tangent vectors contracted with the stress-energy tensor gives $\kappa^\mu \kappa^\nu T_{\mu\nu}=(k^0)^2(\rho+p)$ 
where $\rho$ and $p$ are density and pressure respectively.

Thus the defocusing condition for an FRW metric near the bounce fulfilling $\Box R=r_1 R + r_2$ is
\begin{widetext}
\ba \label{eq:defocusingconditionfinalforFRW}       
        \frac{1}{M^2_P +4 F( r_1) (R+\frac{r_2}{r_1}) -4
f_0\frac{r_2}{r_1}} \bigg[ 2F\left(r_1\right) \left( r_1 R+r_2\right) -F'(r_1)\left(\partial_0 R\right)^2 -(\rho+p)
        \bigg]>0.
\ea
\end{widetext}
where $f_0$ is the zeroeth order coefficient of $F(r_1)=F(\Box=r_1)$. Note that \eqref{eq:defocusingconditionfinalforFRW}
does not depend explicitly on the curvature $\kappa$, which is encoded into the Ricci scalar $R(t)$.

\section{Simplifying our condition}
We can simplify the defocusing condition by placing conditions on $F(\Box)$ and $F'(\Box)$ by inserting a simple solution to the equations of motion. 
\subsection{Inserting the equations of motion}

The vacuum trace equation for metrics satisfying the ansatz is~\cite{Biswas:2010zk,Koshelev:2012qn,Koshelev:2013lfm}
\ba \label{eq:traceequationforansatz}
        A_1 R + A_2 \left(2 r_1 R^2 +( \partial_\mu R)  \partial^\mu R \right) + A_3 = 0, 
\ea
where the coefficients $A_i$ are given by 
\ba
        A_1 &=& -M^2_P + 4 F'(r_1) r_2 - 2 \frac{r_2}{r_1} \left(F(r_1) -f_0\right) + 6 F(r_1)r_1, \non\\
         A_2 &=&  F'(r_1),\non\\ 
        A_3 &=& 4 \Lambda + \frac{r_2}{r_1}\left( M^2_P+ A_1\right) - 2 \frac{r_2^2}{r_1}F'(r_1).        
\ea
The simplest solution to \eqref{eq:traceequationforansatz} is to assume $A_i=0$ which gives the identities
\ba \label{eq:simplesolutiontoeoms}
         F'(r_1)&=&0, \quad \quad r_2 = - \frac{r_1\left(M^2_P - 6 F(r_1)r_1\right)}{2\left(F(r_1)-f_0\right)},\non\\
        \Lambda &=&-\frac{r_2M^2_P}{4r_1}= M^2_P \frac{\left(M^2_p-6 F(r_1)r_1\right)}{8\left(F(r_1)-f_0\right)},
\ea       
which allows us to remove $f_0$ from our defocusing condition \eqref{eq:defocusingconditionfinalforFRW}, which becomes 
\ba \label{eq:defocusingconditionfinalforFRWsimplesoln}       
        \frac{ 2F\left(r_1\right) \left( r_1 R+r_2\right)-(\rho+p)}{ F( r_1)
(4R+6r_1)-M^2_P } >0,
\ea
where again the spatial curvature $\kappa$ does not feature explicitly, but is encoded within $R$. 

In the early universe, we expect the curvature to be very large and we can assume $r_1,r_2\ll R$, so
\ba \label{eq:defocusingconditionfinalforFRWsimplesollargecurvature}    
        \frac{ 2r_1RF\left(r_1\right) -(\rho+p)}{ 4R F( r_1)-M^2_P } >0.
\ea

There are two cases where the defocusing condition is fulfilled
\begin{enumerate}
\item
$2r_1RF\left(r_1\right) -(\rho+p)>0$ and $4R F( r_1)-M^2_P>0$
\item
$2r_1RF\left(r_1\right) -(\rho+p)<0$ and $4R F( r_1)-M^2_P<0$

\end{enumerate}
Note that if $\rho+p$ is very large, as it would be at the bounce point, 
then it is unsatisfactory (but possible) 
for this to be cancelled out by an even larger curvature term. 
However, we do note that it would be natural for the 
curvature to be large at this time. It is indeed pleasing that in the second case, having $r_1>0$ and $RF(r_1)<0$ means that the condition is fulfilled and the singularity is avoided. 

We have shown that it is possible to not only avoid the singularity in the minimal case where 
we ignore matter, but actually for the matter to aid in the defocusing!
We now look at examples where this can apply.

\section{Specific bouncing models}      
\subsection{Toy model}
If we take a simplistic flat bouncing model with the scale factor
\ba
        a(t)=1+a_2t^2,      
\ea
which fulfills
the ansatz with $r_1=-6a_2$      
and $r_2 =48a_2^2$ and
was investigated in~\cite{Conroy:2014dja}.

The defocusing condition for the simple solution to the equations of motion \eqref{eq:defocusingconditionfinalforFRWsimplesoln}
at small times and neglecting matter becomes 
\ba \label{eq:defocusingconditionfinalforFRWsimplesolntoy}       
        \frac{ M^2_P+16a_2f_0-\frac{5}{12}(\rho+p)}{M^2_P +96
f_0a_2} >0
\ea 
which is fulfilled for $-\frac{M^2_P}{16}<a_2f_0<-\frac{M^2_P}{96}$.
\subsection{Assuming a cosh solution}
In this section we assume the solution $a(t)=\cosh(\sigma t)$, where $\sigma$ is a constant. This  vacuum solution to the 
equations of motion was investigated in~\cite{Biswas:2005qr,Biswas:2010zk,Koshelev:2013lfm} 
and return to the de Sitter metric at late times. 
This solution requires a cosmological constant and radiation.
This metric fulfills the ansatz~\footnote{Note that $r_1$ and $r_2$ have opposite signs for both
 the simple solution $1+t^2$ and the cosh solution. Therefore by \eqref{eq:simplesolutiontoeoms}, this is equivalent 
 to a positive $\Lambda$, whereas for the exponential bouncing solution, a negative $\Lambda$ is produced.}
$\Box R=r_1 R +r_2$ with $r_1=2\sigma^2$ and $r_2=-24\sigma^4$. 

Therefore the defocusing condition \eqref{eq:defocusingconditionfinalforFRW} at $t\approx 0$ near the bounce point
for an FRW metric with the scale factor $a(t)=\cosh(\sigma t)$ and non-zero spatial curvature $\kappa$ 
is\footnote{Note that \eqref{eq:defocusingfrwcoshfdashvanishedwithsecondconditionsmalltimes} 
matches \eqref{eq:defocusingconditionfinalforFRW} when we take $r_1 =2\sigma^2$ and $r_2=-24\sigma^4$.} 
\ba \label{eq:defocusingfrwcoshfdashvanishedwithsecondconditionsmalltimes}
        \frac{24\sigma ^2  \left(\kappa-\sigma ^2 \right)F\left(r_1\right) -(\rho+p)}{M^2_P +24 F(2 \sigma^2) 
        \left(\kappa-\sigma ^2 \right) +48\sigma^2 f_0}   <0.
\ea 
Taking the simple solution \eqref{eq:simplesolutiontoeoms} to the equations of motion and neglecting matter, \eqref{eq:defocusingfrwcoshfdashvanishedwithsecondconditionsmalltimes}
is fulfilled for zero spatial curvature $\kappa$ if $-\frac{1}{24}\frac{M^2_P}{\sigma^2}<f_0<-\frac{1}{96}\frac{M^2_P}{\sigma^2}$.
If matter is included, and gives a very large contribution, then defocusing occurs for $f_0<-\frac{1}{96}\frac{M^2_P}{\sigma^2}$. 

We have shown that it is possible for a bouncing cosh solution, which was already shown to avoid the Big Bang curvature singularity, 
to avoid the Hawking-Penrose singularity for certain curvature. 
\subsection{Exponential bouncing solution}
Finally we look at a flat FRW metric with the exponential bouncing scale factor which was studied in~\cite{Koshelev:2012qn,Koshelev:2013lfm}
where 
\ba \label{eq:exponentialbouncingsoln}
        a(t)=e^{\frac{\lambda}{2} t^2}.
\ea
where $\lambda$ is a positive constant with dimensions of mass squared. 
This scale factor gives the Ricci scalar $R=3\lambda\left(1+\lambda t^2\right)$ which fulfills the ansatz with $\Box R(t) = -6\lambda R(t) -12\lambda^2$.
The defocusing condition \eqref{eq:defocusingconditionfinalforFRW} for small times becomes
\ba \label{eq:defocusingconditionfinalforFRWsimplesolnexpbouncingnearbounce}       
        \frac{ 5\lambda^2F\left(r_1\right)+(\rho+p)}{M^2_P +20\lambda F( r_1)  -8
f_0\lambda} <0.
\ea
and with the simple solution to the equations of motion \eqref{eq:simplesolutiontoeoms}, 
we find that there is defocusing for $-\frac{1}{4}M^2_P<\lambda f_0 <\frac{1}{28}M^2_P$ if we neglect matter, 
or $\lambda f_0 <\frac{1}{28}M^2_P$ if the contribution of matter is large.

The exponential bouncing solution \eqref{eq:exponentialbouncingsoln} generated using ghost-free Infinite Derivative Gravity
was already known to avoid the Big Bang curvature singularity, 
but we have shown it can avoid the Hawking-Penrose singularity, and that the addition of matter can actually aid this process. 

\section{Conclusion}  
We have shown that IDG can allow defocusing of null rays, and thus avoid Hawking-Penrose singularities, 
for more general spacetimes as long
as our simplifying ansatz holds. We first looked at completely general spacetimes where this ansatz applies, 
before restricting ourselves to time-dependent Ricci scalars and then time-dependent metrics. 
We looked at an FRW metric and showed that this could defocus given certain conditions. 

We investigated three specific scale factors within an FRW metric. 
We first looked at a simple polynomial bouncing universe, then a cosh scale factor and finally
an example of a exponentially bouncing time-dependent metric. We showed that all of these metrics 
allowed defocusing in the minimal case, but they also allowed defocusing in the case where matter gave a large 
contribution, as in the early universe.

It has previously been shown that IDG would permit a bouncing universe which avoided a curvature singularity. We have now shown that
it is also possible to avoid the Hawking-Penrose singularity in these universes.

\section{Acknowledgements}
We would particularly like to thank Anupam Mazumdar for his guidance and suggestions on this project. We would also like to thank Alexey Koshelev and Aindriu Conroy for their help on this work.


\end{document}